\documentstyle[12pt]{article}
\begin{document}
\def\PRL{Phys.~Rev.~Lett.~}
\def\PRD{Phys.~Rev.~D}
\def\NPB{Nucl.~Phys.~B}
\def\PL{Phys.~Lett.~}
\def\PLB{Phys.~Lett.~B}
\def\Slash{\hskip -.6em/}
\begin{flushright}
EFI 93-40 \\
FERMILAB-PUB-93/247-T \\
hep-ph/9309307 \\
September 1993 \\
\end{flushright}
\bigskip
\medskip
\begin{center}
\large
{\bf Oblique Corrections to the W Width} \\
\bigskip
\medskip
\normalsize
{\it Jonathan L. Rosner and Mihir P. Worah} \\
\medskip
{\it Enrico Fermi Institute and Department of Physics \\
University of Chicago \\
5640 S. Ellis Avenue, Chicago, IL 60637} \\
\bigskip

and
\bigskip

{\it Tatsu Takeuchi \\
\medskip
Fermi National Accelerator Laboratory \\
P.O. Box 500, Batavia, IL 60510 } \\
\bigskip
\bigskip
{\bf ABSTRACT} \\
\end{center}
\begin{quote}
The lowest-order expression for the partial $W$ width to $e \nu ,~\Gamma (W \to
e \nu) = G_\mu M_W^3 /(6 \pi \sqrt{2})$, has no oblique radiative corrections
from new physics if the measured $W$ mass is used.  Here $G_\mu = (1.16639 \pm
0.00002) \times 10^{-5}$ GeV/$c^2$ is the muon decay constant.  For the present
value of $M_W = (80.14 \pm 0.27)$ GeV/$c^2$, and with $m_t = 140$ GeV$/c^2$,
one expects $\Gamma (W \to e \nu) = (224.4 \pm 2.3)$ MeV. The total width
$\Gamma_{\rm tot}(W)$ is also expected to lack oblique corrections from new
physics, so that $\Gamma_{\rm tot} (W)/ \Gamma (W \to e \nu ) = 3 + 6 [1 + \{
\alpha_s (M_W)/\pi \}]$. Present data are consistent with this prediction.
\end{quote}
\newpage

\centerline{\bf I.  INTRODUCTION}
\bigskip
Precise measurements of electroweak phenomena have reached a level of accuracy
which permits the search for new phenomena, manifesting themselves through
radiative corrections. A particularly interesting class of such effects occur
through loops of new particles in $W$ and $Z$ propagators, and are known as
``oblique'' corrections [1].

The effects of oblique corrections have been studied in the past few years by
several groups [2-6]. By expanding vacuum polarization tensors for $\gamma -
\gamma , ~ \gamma - Z , ~ Z -Z$, and $W-W$ self-energies to order $q^2
/M_{\rm new}^2$ where $M_{\rm new}$ is the mass scale associated with new
physics, one can express electroweak observables as nominal values (for a
specific mass of the top quark and Higgs particle) corrected by linear
functions of a few phenomenological variables. These variables encapsulate the
effects of new physics on the observables in a concise way. Thus, for example,
in the notation of Ref.~[5], one has variables $S_W , S_Z$, and $T$, where
$S_W$ and $S_Z$ describe the effects linear in $q^2$ of $W$ and $Z$ wave
function renormalization due to new particles, while $T$ is sensitive to
violations of custodial SU(2) [7] such as occur in the case of a very heavy top
quark.

In the present paper we shall show that when the $W$ partial width to $e \nu$
and total width are expressed in terms of the measured muon decay constant
$G_\mu = (1.16639 \pm 0.00002) \times 10^{-5}~ {\rm GeV}^{-2}$ and $W$ mass
$M_W = 80.14 \pm 0.27$ GeV (the average of values from Refs.~[8] and [9]), the
lowest-order expressions do not receive corrections proportional to $S_W ,
S_Z$, or $T$.  The relative smallness of standard model corrections to the $W$
partial and total widths when expressed in this manner has been noted in
Refs.~[10] and [11].  A recent treatment of the $W$ width in the context of
such parameters has appeared in Ref.~[12], but the result mentioned here does
not appear explicitly.

The predicted partial and total widths are:
\begin{equation}
\Gamma (W \to e \nu) =
\frac{G_\mu M_W^3}{6 \pi \sqrt{2}} [ 1 + \delta^{\rm sm}]
= (224.4 \pm 2.3) ~ {\rm MeV}~~~,
\end{equation}
\begin{equation}
\Gamma_{\rm tot} (W) = \{3 + 6 [1 + \alpha_s (M_W) / \pi ] \} \Gamma (W \to e
\nu ) = (2.07 \pm 0.02 ) ~ {\rm GeV}~~~,
\end{equation}
where most of the errors come from that on $M_W$, and $\delta^{\rm sm}$ is a
small correction in the standard model, whose value [11] is about $ - 0.35 \%$
when evaluated for the nominal values $m_t = 140$ GeV/$c^2$ and $M_H = 100$
GeV/$c^2$.

Most standard model corrections have already been absorbed into $G_\mu$ and/or
the physical value of $M_W$, which explains why $\delta^{\rm sm}$ is only
a few parts in $10^3$. Consequently, a precise measurement of $\Gamma_{\rm
tot} (W)$ (to a level of 1\%) would begin to check $M_W$ itself at levels
comparable to present direct measurements. Deviations from the predictions (1)
and (2) would indicate physics outside the purview of the parameters $S_W ,
S_Z$, and $T$. We shall mention such possibilities at the end of this article.

Our discussion is organized as follows. In Section II we introduce $S_W , S_Z$,
and $T$, and show that the expressions (1) and (2) do not receive corrections
linear in these parameters.  In Section III we discuss the full set of standard
model corrections. In Section IV we present details leading to the numerical
values in (1) and (2), and compare these results with recent experiments. In
Section V we note the role of corrections of higher order in $q^2/M_{\rm
new}^2$ which have recently been mentioned in [12] (as well as the earlier
discussion of Ref.~[13]).  We cite possible sources of deviation from the
predictions (1) and (2). A suggestion for measuring the absolute $W$ width
using continuum production of lepton pairs is noted in Section VI, while
Section VII summarizes.  Explicit formulae involving top quark and Higgs boson
contributions to corrections to the $W$ width are noted in an Appendix.
\bigskip

\centerline{\bf II.  ABSENCE OF NEW-PHYSICS OBLIQUE CORRECTIONS}
\bigskip

In this section, we will first introduce the oblique correction parameters
$S_W$, $S_Z$, and $T$ and then show that the prediction for the $W$ width is
independent of these parameters when the muon decay constant,
 $G_\mu$, and the $W$ mass, $M_W$, are used as inputs.

When considering oblique corrections in the $SU(2)_L \times U(1)_Y$ gauge
theory of electroweak interactions, there are four types of vacuum
polarizations that must be taken into account.  They are the self-energies of
the photon, the $Z$, and the $W$, and the $Z$--photon mixing, which we denote
$\Pi_{AA}(q^2)$, $\Pi_{ZZ}(q^2)$, $\Pi_{WW}(q^2)$ and $\Pi_{ZA}(q^2)$,
respectively [14].  We divide these vacuum polarizations into two parts:
\begin{equation}
\Pi_{XY}(q^2) = \Pi_{XY}^{\rm sm}(q^2) + \Pi_{XY}^{\rm new}(q^2)
\end{equation}
for $(XY) = (AA), (ZA), (ZZ), (WW)$, where $\Pi_{XY}^{\rm sm}(q^2)$ is the
contribution of the standard model, and $\Pi_{XY}^{\rm new}(q^2)$ is the
contribution of new physics. If we assume the scale of new physics, $M_{\rm
new}$, which contributes to the $\Pi_{XY}^{\rm new}$'s to be large compared to
the $W$ and $Z$ masses, it is then reasonable to expand the new physics
contributions around $q^2=0$ and neglect higher orders which will be suppressed
by powers of $q^2/M_{\rm new}^2$.  Keeping terms linear in $q^2$, we find
\begin{eqnarray}
\Pi_{AA}^{\rm new}(q^2) & = & q^2 \Pi_{AA}^{\prime\;\rm new}(0) + \cdots
\nonumber\\
\Pi_{ZA}^{\rm new}(q^2) & = & q^2 \Pi_{ZA}^{\prime\;\rm new}(0) + \cdots
\nonumber\\
\Pi_{ZZ}^{\rm new}(q^2) & = & \Pi_{ZZ}^{\rm new}(0)
                        + q^2 \Pi_{ZZ}^{\prime\;\rm new}(0) + \cdots
\nonumber\\
\Pi_{WW}^{\rm new}(q^2) & = & \Pi_{WW}^{\rm new}(0)
                        + q^2 \Pi_{WW}^{\prime\;\rm new}(0) + \cdots
\nonumber\\
\label{PiTaylor}
\end{eqnarray}
Note that $\Pi_{AA}^{\rm new}(0) = \Pi_{ZA}^{\rm new}(0) = 0$ from QED gauge
invariance. Thus, in this approximation, the contribution of new physics can be
parametrized by just six numbers: $\Pi_{AA}^{\prime\;\rm new}(0)$,
$\Pi_{ZA}^{\prime\;\rm new}(0)$, $\Pi_{ZZ}^{\rm new}(0)$,
$\Pi_{ZZ}^{\prime\;\rm new}(0)$, $\Pi_{WW}^{\rm new}(0)$, and
$\Pi_{WW}^{\prime\;\rm new}(0)$. Three linear combinations of these numbers
will be absorbed into the renormalization of the three input parameters used to
fix the theory. That will leave us with only three linear combinations that are
finite and observable. A popular choice for the three combinations is [5]
\begin{eqnarray}
\alpha S_Z & = & 4s^2c^2
                 \left[ \Pi_{ZZ}^{\prime\;\rm new}(0)
                      - \frac{ c^2 - s^2 }{ sc }\,\Pi_{ZA}^{\prime\;\rm new}(0)
                      - \Pi_{AA}^{\prime\;\rm new}(0)
                 \right]    \\
\alpha S_W & = & 4s^2
                 \left[ \Pi_{WW}^{\prime\;\rm new}(0)
                      - \frac{ c }{ s }\,\Pi_{ZA}^{\prime\;\rm new}(0)
                      - \Pi_{AA}^{\prime\;\rm new}(0)
                 \right]    \\
\alpha T   & = & \frac{ \Pi_{WW}^{\rm new}(0) }{ M_W^2 }
               - \frac{ \Pi_{ZZ}^{\rm new}(0) }{ M_Z^2 },
\end{eqnarray}
where
\begin{equation}
c = \frac{ g }{ \sqrt{ g^2 + g^{\prime 2} } }, \qquad
s = \frac{ g'}{ \sqrt{ g^2 + g^{\prime 2} } }.
\end{equation}
In the notation of Ref.~[2], $S_Z = S$, and $S_W = S+U$.

The effect of oblique corrections from new physics to an observable, ${\cal
O}$, can be expressed in terms of the the parameters $S_Z$, $S_W$, and $T$ as
\begin{equation}
{\cal O}_{\rm th} = {\cal O}_{\rm sm} [ 1 + a S_W + b S_Z + c T ]
\end{equation}
where ${\cal O}_{\rm sm}$ is the Standard Model prediction while ${\cal O}_{\rm
th}$ is the theoretical prediction including oblique corrections from new
physics. The coefficients $a$, $b$, and $c$ depend on the observable ${\cal O}$
and are easily calculable. Now, an important point which is not often mentioned
explicitly is that {\it both the Standard Model prediction ${\cal O}_{\rm sm}$
and the coefficients $a$, $b$, $c$ depend on which three observables are used
as inputs to fix the theory.} To give a trivial example, consider using
$\alpha$, $G_\mu$, and $M_Z$ as inputs to predict $M_W$. In this case, the
theoretical prediction for $M_W$ will be given by
\begin{equation}
M_{W, \rm th}^2 = M_{W, \rm sm}^2(\alpha, G_\mu, M_Z)
        \left[ 1 + \frac{ \alpha }{c^2 - s^2}
                   \left( \frac{ c^2 - s^2 }{ 4s^2 }S_W
                         -\frac{ 1 }{ 4s^2 }S_Z
                         + c^2 T
                   \right)
        \right].
\end{equation}
However, if the value of $M_W$ itself is used as one of the three inputs,
then the theoretical ``prediction'' will be
\begin{equation}
M_{W,\rm th}^2 = M_{W,\rm sm}^2(M_W,*,*) = M_W^2
\end{equation}
and there will be no extra corrections from $S_W$, $S_Z$, or $T$.

The observation that we would like to make in this paper is that if
$G_\mu$ and $M_W$ are used as inputs to predict the $W$ width, $\Gamma_W$,
then $\Gamma_W$ doesn't receive any extra corrections from $S_W$, $S_Z$,
and $T$.  Thus
\begin{equation}
\Gamma_{W,\rm th} = \Gamma_{W,\rm sm}(G_\mu,M_W,*).
\end{equation}
This is for the simple reason that $\Gamma_W$ receives corrections from new
physics through the two parameters $\Pi_{WW}^{\rm new}(0)$ and
$\Pi_{WW}^{\prime\;\rm new}(0)$, but these happen to be the ones that are
absorbed into the renormalizations of $G_\mu$ and $M_W$ and are unobservable.
We
will show this more explicitly in the following.

Consider the obliquely corrected $W$ propagator:
\begin{equation}
G_{WW}(q^2)
= \frac{ 1 }{\displaystyle  q^2 - \frac{ g^2v^2 }{ 4 } - \Pi_{WW}(q^2) }
\end{equation}
where $g^2v^2/4$ is the bare $W$ mass.  If we rewrite this propagator in
terms of the physical $W$ mass
\begin{equation}
M_W^2 = \frac{ g^2v^2 }{ 4 } + \Pi_{WW}(M_W^2)
\label{MWdef}
\end{equation}
and the wave function renormalization constant [15]
\begin{equation}
Z_W^{-1} = 1 - \Pi'_{WW}(M_W^2)
\end{equation}
we find
\begin{equation}
G_{WW}(q^2)
= \left[ \frac{ 1 }{ 1 + \delta_W(q^2) } \right]
  \left( \frac{ Z_W }{ q^2 - M_W^2 } \right)
\label{GWW}
\end{equation}
where
\begin{equation}
\delta_W(q^2) \equiv
             Z_W  \left[ \Pi'_{WW}(M_W^2)
                          - \frac{ \Pi_{WW}(q^2) - \Pi_{WW}(M_W^2) }
                                 { q^2 - M_W^2 }
                  \right].
\end{equation}
Note that $\delta_W(M_W^2)=0$. Now, since
\begin{equation}
-\frac{ 4 G_\mu }{ \sqrt{2} } = \frac{ g^2 }{ 2 }G_{WW}(0)
\end{equation}
(up to certain vertex and box corrections from muon decay which will be
discussed in Sec.~III),
Eq.~(\ref{GWW}) leads to
\begin{equation}
g^2 Z_W =  4\sqrt{2}G_\mu M_W^2 [ 1 + \delta_W(0) ]~~~.
\label{gZWGF}
\end{equation}
Using this result, the partial width of the decay $W \rightarrow e\nu$ can be
written as
\begin{equation}
\Gamma(W\rightarrow e\nu)_{\rm th}
 = \frac{ g^2 M_W }{ 48\pi }Z_W
 = \frac{ G_\mu M_W^3 }{ 6\pi\sqrt{2} }[ 1 + \delta_W(0) ]~~~,
\end{equation}
where the effect of oblique corrections is summarized in $\delta_W(0)$.

Separating $\delta_W(0)$ into the standard model contribution, $\delta_W^{\rm
sm}(0)$, and the contribution of new physics, $\delta_W^{\rm new}(0)$, we find
\begin{eqnarray}
\Gamma(W\rightarrow e\nu)_{\rm th}
& = & \frac{ G_\mu M_W^3 }{ 6\pi\sqrt{2} }
      [ 1 + \delta_W^{\rm sm}(0) + \delta_W^{\rm new}(0) ] \nonumber\\
& = & \frac{ G_\mu M_W^3 }{ 6\pi\sqrt{2} }
      [ 1 + \delta_W^{\rm sm}(0)][ 1 + \delta_W^{\rm new}(0) ]  \nonumber\\
& \equiv & \Gamma(W\rightarrow e\nu)_{\rm sm} [ 1 + \delta_W^{\rm new}(0) ]
\label{GammaW}
\end{eqnarray}
Now if we Taylor expand $\Pi_{WW}^{\rm new}(q^2)$ in the definition of
$\delta_W^{\rm new}(q^2)$, we find
\begin{equation}
\delta_W^{\rm new}(0) = \frac{ M_W^2 }{ 2 }\Pi_{WW}^{\prime\prime\;\rm new}(0)
                  + \cdots
\end{equation}
which shows explicitly that $\Pi_{WW}^{\rm new}(0)$ and $\Pi_{WW}^{\prime\:\rm
new}(0)$ disappear from Eq.~(\ref{GammaW}); they have been absorbed into
$G_\mu$ and $M_W$ through Eqs.~(\ref{MWdef}) and (\ref{gZWGF}). Therefore, in
the approximation where the $\Pi_{XY}^{\rm new}(q^2)$'s are expanded only up to
the linear term in $q^2$, $\delta_W^{\rm new}(0)$ can be safely neglected.

An exactly analogous argument can show that
\begin{equation}
\Gamma(Z\rightarrow \nu\bar{\nu})_{\rm th}
= \frac{ G_\mu M_Z^3 }{ 12\pi\sqrt{2} }\rho [ 1 + \delta_Z(0) ]
\end{equation}
where
\begin{equation}
\delta_Z(q^2) \equiv Z_Z
         \left[ \Pi'_{ZZ}(M_Z^2)
                - \frac{ \Pi_{ZZ}(q^2) - \Pi_{ZZ}(M_Z^2) }
                       { q^2 - M_Z^2 }
         \right].
\end{equation}
Again, there will be no $S_W$, $S_Z$, or $T$ dependence coming from
$\delta_Z(0)$. However, $\Gamma(Z\rightarrow \nu\bar{\nu})$ will receive $T$
dependence through the $\rho$--parameter:
\begin{eqnarray}
\rho_{\rm th} & = &  1 + \delta\rho_{\rm sm} + \alpha T          \nonumber \\
          & = & (1 + \delta\rho_{\rm sm} )( 1 + \alpha T )   \nonumber \\
          & = & \rho_{\rm sm} ( 1 + \alpha T )
\end{eqnarray}
Therefore, writing $\delta_Z(0) = \delta_Z^{\rm sm}(0) + \delta_Z^{\rm new}(0)$
and neglecting $\delta_Z^{\rm new}(0)$, we find
\begin{equation}
\Gamma(Z\rightarrow \nu\bar{\nu})_{\rm th}
= \Gamma(Z\rightarrow \nu\bar{\nu})_{\rm sm}(1 + \alpha T)
\end{equation}
where
\begin{equation}
\Gamma(Z\rightarrow \nu\bar{\nu} )_{\rm sm}
\equiv  \frac{ G_\mu M_Z^3 }{ 12\pi\sqrt{2} }
        \rho_{\rm sm} [ 1 + \delta_Z^{\rm sm}(0) ],
\end{equation}
so that a measurement of the partial width (given the precise value $M_Z =
(91.187 \pm 0.007) $ MeV obtained at LEP [16]) provides information on $T$.

The total width of the $W^+$ is calculated under the assumption that the open
decay channels are $e^+ \nu_e , \mu^+ \nu_\mu , \tau^+ \nu_\tau$, and three
colors of $u \bar{d}$ and $c \bar{s}$. Fermion masses (treated in [10] and
[11])
give negligible effects, reducing the total predicted $W$ width by less than 1
MeV. Thus we obtain the expression (2), where the factor of $1+  \alpha_s
(M_W)/\pi$ is the usual QCD correction [17] for decays into colored quarks. The
expression (2), like (1), does not have any correction factors involving $S_W ,
S_Z$, or $T$.

In a treatment where the terms up to those that are quadratic in $q^2$ are kept
in Eq.~(\ref{PiTaylor}), $\delta_W^{\rm new}(0)$ and $\delta_Z^{\rm new}(0)$
cannot be neglected. In Ref.~[12], Maksymyk, Burgess, and London use the
notation
\begin{equation}
\alpha V \equiv \delta_Z^{\rm new}(0),\qquad
\alpha W \equiv \delta_W^{\rm new}(0).
\end{equation}
and discuss the possible sizes of $V$ and $W$. In their notation,
\begin{eqnarray}
\frac{ \Gamma(W\rightarrow e\nu)_{\rm th} }
     { \Gamma(W\rightarrow e\nu)_{\rm sm} }
& = & 1 + \alpha W~~~,    \\
\frac{ \Gamma(Z\rightarrow \nu\bar{\nu})_{\rm th} }
     { \Gamma(Z\rightarrow \nu\bar{\nu})_{\rm sm} }
& = & 1 + \alpha T + \alpha V~~~.
\end{eqnarray}
We shall comment on possible sources of $W$ in Sec.~V.
%
\bigskip

\centerline{\bf III.  FULL SET OF STANDARD MODEL CORRECTIONS}
\bigskip

As mentioned above, the tree level expression for the partial $W$ width,
\begin{equation}
\Gamma(W\rightarrow e\nu) = {G_\mu M_{W}^3\over{6\pi\sqrt{2}}},
\end{equation}
accounts for most of the leading order standard model oblique corrections, as
well as the ``new'' oblique corrections, parametrized by $S_W,~ S_Z$, and $T$.
The oblique corrections not absorbed into $G_\mu$ and $M_W$ are given by
\begin{equation}
\delta_W(0) = Z_W \left[ \Pi_{WW}'(M_{W}^2) -
{\Pi_{WW}(M_{W}^2)-\Pi_{WW}(0)\over{M_{W}^2}} \right] ~~~.
\end{equation}
The complete corrected result will be given by
\begin{equation}
\Gamma(W\rightarrow e\nu) = {G_\mu M_W^3 \over 6 \pi \sqrt 2} \left[ 1 +
\delta_W^{\rm sm}(0) + \delta_V^{\rm sm} + \delta_\mu \right]~~~,
\end{equation}
where $\delta_V^{\rm sm}$ expresses the effect of the vertex and bremsstrahlung
[11,18] corrections, and
\begin{equation}
\delta_\mu = - \frac{G_\mu M_W^2}{2 \pi^2 \sqrt 2}
\left[ 4 \left( \Delta - \ln\frac{ M_W^2 }{ \mu^2 } \right)
     + \left( 6 + \frac{7 - 4 s^2}{2 s^2} \ln c^2 \right) \right]~~~,
\end{equation}
with
\begin{equation}
\Delta \equiv {1\over \epsilon} - \gamma_E + \ln 4\pi
\end{equation}
takes care of the vertex and box corrections specific to muon decay
which have been omitted in Eq.~(18) [19].
Note that $\delta_W^{\rm sm}(0)$ in Eq.~(33) is UV finite, while the
UV divergences in $\delta_V^{\rm sm}$ and $\delta_\mu$ cancel against
each other.
However, there is an IR divergence in $\delta_W^{\rm sm}(0)$ coming
from the $\gamma-W$ loop, which is cancelled by a similar divergence
in $\delta_V^{\rm sm}$.
The finite contributions to
$\delta^{\rm sm} = \delta_W^{\rm sm}(0) + \delta_V^{\rm sm} + \delta_\mu$
 are summarized in Tables I and II for
$m_t = 140$ GeV/$c^2$ and $M_H = 100$ GeV/$c^2$,
with $g^2 Z_W/(4 \pi)^2 = G_\mu
M_W^2 / (2 \pi^2 \sqrt 2) = 0.268 \%,~s^2 = 0.23$.
\bigskip

Putting all the standard model corrections together, we find that the standard
model correction to Eq.~(1) is $\delta^{\rm sm} = - 0.35 \%$.  The difference
between the correction for leptons and for quarks is too small to affect the
ratio (2) appreciably.
\bigskip

\centerline{\bf IV.  NUMERICAL EVALUATION}
\bigskip
The two most precise estimates of the $W$ mass come from the CDF and UA2
collaborations:
\begin{equation}
M_W ({\rm measured}) =
\left \{
\begin{array}{l}
79.92 \pm 0.39 ~{\rm GeV}/c^2 ~~ [8] \\
\\
80.35 \pm 0.37 ~ {\rm GeV}/c^2 ~~ [9] \\
\\
80.14 \pm 0.27 ~ {\rm GeV}/c^2 ~~ ({\rm average}), \\
\end{array}
\right .
\end{equation}
where we have recalibrated the UA2 value [9] in terms of the known $Z$
mass. For $\alpha_s (M_W)$ we use an error attributed to systematic differences
among various determinations [20], and take $\alpha_s (M_W) = 0.12 \pm 0.01$.

Two recent determinations of the $W \to e \nu$ branching ratio have been
performed [21,22]. The method [23] relies upon the measurement of
\begin{eqnarray}
\label{rat}
\frac{\sigma (\bar{p} p \to e^\pm \nu + \ldots )}{\sigma ( \bar{p} p \to
e^+ e^- + \ldots )} & = &
\frac{\sigma ( \bar{p}p \to W^\pm + \ldots )}{\sigma ( \bar{p}p \to Z +
\dots )}
\frac{\Gamma_{\rm tot} (Z)}{\Gamma (Z \to e^+ e^-)} \nonumber \\
& \times & \frac{\Gamma (W^+ \to e^+ \nu )}{\Gamma_{\rm tot} (W)} ~~~.
\end{eqnarray}
The measured values of the left\--hand side are $10.64 \pm 0.36 \pm 0.27$
(Ref.~[21]), $10.0 \pm 1.1 \pm 2.4$ (muon channels, Ref.~[22]), and $10.56 \pm
0.87 \pm 1.07$ (electron channels, Ref.~[22]). The first ratio on the
right\--hand side is taken from theory to be $3.23 \pm 0.03$ [24] (CDF) or
$3.26 \pm 0.08$ [25] (D0).
The ratio $\Gamma_{\rm tot} (Z) / \Gamma (Z \to e^+e^-)$
is found from LEP averages [26] to be $29.69 \pm 0.13$. Here we have used
$\Gamma_{\rm tot} (Z) = (2.489 \pm 0.007)$ GeV, $\Gamma (Z \to e^+e^-) = (83.82
\pm 0.27)$ MeV.

The results are
\begin{equation}
\frac{\Gamma (W^+ \to e^+ \nu)}{\Gamma_{\rm tot} (W)} =
\left \{
\begin{array}{l}
0.111 \pm 0.005 ~~ [22] ~~~,\\
\\
0.108 \pm 0.013 ~~ [23] ~~~.\\
\end{array}
\right .
\end{equation}
This is to be compared with the theoretical estimate, made assuming the open
decay channels are $e \nu , \mu \nu , \tau \nu$, $u \bar{d} $, and $c \bar{s}$:
\begin{equation}
\frac{\Gamma (W^+ \to e^+ \nu )}{\Gamma_{\rm tot} (W)} =
[3+6 (1 + \frac{\alpha_s (M_W)}{\pi} ) ]^{-1} = 0.1084 \pm 0.0002~~~.
\end{equation}
The measurement of this ratio does {\em not} test $\Gamma (W \to e^+ \nu)$ or
$\Gamma_{\rm tot}(W)$ separately.

The small difference between the standard model corrections for quark and
lepton final states leads to an increase of the above ratio by about $3 \times
10^{-5}$, or 0.03\% of its value.
\bigskip

\centerline{\bf V.  POSSIBLE SOURCES OF DEVIATION}
\bigskip

The partial width $\Gamma (W \to e \nu )$ could be affected by mixing of the
$W$ with other states (e.g., new gauge bosons or vector mesons in the TeV
region associated with substructure of the Higgs sector [27]. We expect,
however, that constraints from other data would severely limit such mixing.

The ratio $\Gamma_{\rm tot} (W)/ \Gamma (W \to e \nu )$ could be raised from
its predicted value if additional exotic decay channels for the $W$ were
available. Such a channel could be $t + \bar{b}$, where the $t$ decays to a
charged Higgs boson and a $b$ quark. The result of Ref.~[21] implies $m_t > 62$
GeV under such a scenario. Another such channel would be a pair of scalar
bosons $H^+ H^0$. Comparison of the predicted and observed branching ratios
places severe limits on the couplings for such decays.

As an example of the effects [12] due to higher-order oblique corrections from
``new'' physics, we calculate $\delta_{W}^{\rm new}(0)$ in the
two-Higgs-doublet extension of the standard model [28]. We choose
$m_1=m_2=m_{+}/4=m_3/8$ for the scalar masses, where $m_1$ and $m_2$ are the
masses of the neutral scalars,  $m_{+}$ the charged scalar, and $m_3$ the
neutral pseudoscalar. This choice is of interest since for $m_3 \geq 500$ GeV
one obtains a negative contribution to the parameter $\rho$ [29]. We plot our
results as the dashed line in Fig.~1.

The authors of Ref.~[12] calculate the contribution to $\delta_{W}^{\rm new}
(0)$ ($\alpha W$ in their language) of a doublet of heavy degenerate leptons.
We reproduce this calculation and plot the result as the dotted line in Fig.~1.
Both this result and that of the previous paragraph lead to very small and
probably undetectable effects on the $W$ partial and total widths.

Very recently Lavoura and Li [30] have pointed out that one can increase some
of the parameters introduced in Refs. [12] and [13] without correspondingly
large increases in $S_W$, $S_Z$, and $T$ by introducing scalar multiplets of
very high weak isospin.  However, it appears difficult in the cases they
consider to obtain any detectable changes in the $W$ width without appreciable
effects elsewhere.
\bigskip

\centerline{\bf VI.  MEASUREMENT OF ABSOLUTE WIDTH}
\bigskip

The reaction $\bar p p \to \ell \nu_\ell = \ldots$, where $\ell = e,~\mu,
{}~\tau$, is dominated by the production of real $W$ bosons, but there is
a measurable continuum of events above the $W$ [31,32]. By comparing the
signal for real and virtual $W$ bosons, one can obtain an estimate of
the total width [33].

Let us imagine that partons $i$ and $j$ (typically a $u$ quark and a $\bar d$
antiquark) with squared center-of-mass energy $\hat s$ collide to form
either a real or a virtual $W^+$, which subsequently decays to $\ell^+
\nu_\ell$.  The cross section for this subprocess has the form
\begin{equation}
\frac{d \sigma}{d \hat s} = {\rm Const} \times
\frac{\Gamma_{ij} \Gamma_{\ell \nu_\ell}}
{(\hat s - M_W^2)^2 + M_W^2 \Gamma_{\rm tot}^2}~~~,
\end{equation}
where $\Gamma_{ij}$ is the partial width for the decay of the $W$ into $ij$,
while $\Gamma_{\rm tot}$ is the total $W$ width.  The integral of this cross
section over $\hat s$ is proportional to $\Gamma_{ij} \Gamma_{\ell \nu_\ell} /
\Gamma_{\rm tot}$, while far above $\hat s = M_W^2$ the expression is almost
independent of $\Gamma_{\rm tot}$.  Thus, a comparison of the real $W$ signal
with the continuum $\ell \nu_\ell$ signal above the $W$ normalizes the
production process and gives a measurement of the total $W$ width.

The 1988-9 CDF data [31] indicate that one can count on about four or five
$e^\pm \nu_e$ events above a transverse mass of 100 GeV/$c^2$ for each inverse
pb of integrated luminosity.  Thus, with one inverse femtobarn of data and
detection of both $e^\pm \nu_e$ and $\mu^\pm \nu_\mu$ pairs, one can hope
for a statistical accuracy of about a percent in measurement of $\Gamma_
{\rm tot}$.
\newpage

\centerline{\bf VII.  SUMMARY}
\bigskip

We have shown that lowest-order expression for the $W^+$ partial width to $e^+
\nu_e$ does not receive contributions from new physics contained in the oblique
correction parameters $S_W,~S_Z$, and $T$ when expressed in terms of the muon
decay constant $G_\mu$ and the measured $W$ mass.  As a result, a measurement
of $\Gamma_W$ provides independent information on $M_W$. Any inconsistency
between the value of $M_W$ inferred from the $W$ width and that measured
directly will have to be ascribed to effects not encompassed in these three
parameters.

The present method for measuring $\Gamma_{\rm tot}(W)$ at hadron colliders
actually yields the {\em branching ratio} for $W \to e \nu$. Recent precise
experiments are consistent with the prediction that this ratio should be given
by approximately
\begin{equation}
\frac{\Gamma (W \to e \nu)}{\Gamma_{\rm tot}(W)} = \frac{1}{9}
\left[ 1 + \frac{2}{3} \frac{\alpha_s(M_W)}{\pi} \right]^{-1} ~~~.
\end{equation}
One is still in search of an {\em absolute} measurement of the $W$ partial or
total width. As we have shown, there is not much room for deviations from the
predictions (1) and (2) for these quantities. Comparison of production of real
and virtual $W$ bosons may begin to shed light on the total $W$ width.
\bigskip

\centerline{\bf ACKNOWLEDGMENTS}
\bigskip

We are grateful to Henry Frisch and Sacha Kopp for asking the questions which
led to this investigation, and to David Saltzberg for a study of the
feasibility of the method noted in Sec.~VI.  M.~P.~W. acknowledges useful
conversations with Aaron Grant and Paco Solis. Part of this investigation was
performed while J.~L.~R. was at the Aspen Center for Physics. This work was
supported in part by the United States Department of Energy under Grant Number
DE FG02-90ER40560 and Contract Number DE-AC02-76CH03000.
\bigskip

\centerline{\bf APPENDIX: TOP QUARK AND HIGGS BOSON CONTRIBUTIONS}
\bigskip

The standard model oblique correction due to the $t-\bar b$ loop is
\begin{equation}
\delta_W^t(0) = {g^2 Z_W \over 16\pi^2} \frac{3}{2} \left\{ \frac{2}{3} -
\frac{\xi} {2} - \xi^2 - \xi (1 - \xi^2) \ln[\xi /(\xi - 1)] \right\}~~~,
\end{equation}
with $\xi \equiv m_t^2/M_W^2$.  (We have neglected $m_b$ here.) This quantity
is generally small, and goes to 0 as $m_t\rightarrow \infty$. We plot
$\delta_W^t(0)$ as a function of $m_t$ as the solid line in Fig.~1. For a
nominal top quark mass of 140 GeV, we get
\begin{equation}
\delta_W^t(0) = -{0.15g^2 Z_W \over{16\pi^2}}~~~.
\end{equation}

The standard model Higgs boson's contribution is extremely small:
\begin{eqnarray}
\delta_W^{\rm Higgs}(0) & = & \frac{g^2 Z_W}{4 (4 \pi)^2} \left[ \left(
\frac{47}{6} - \frac{7}{2} \xi_H + \xi_H^2 \right) \right. \nonumber \\
& + & \left. \frac{-4 + 22 \xi_H - 17 \xi_H^2 + 6 \xi_H^3 - \xi_H^4}{2 (\xi_H -
1)} \ln \xi_H \right. \nonumber \\
& + & \left. (-28 + 20 \xi_H - 7 \xi_H^2 + \xi_H^3) \sqrt{ \frac{\xi_H}{4 -
\xi_H}} \arctan \sqrt{ \frac{4 - \xi_H}{\xi_H}} \right] ~~~
\end{eqnarray}
where $\xi_H \equiv m_H^2/m_W^2$.  With $g^2 Z_W \approx 0.4$ and for $M_H =
100$ GeV/$c^2$, we get $\delta_W^{\rm Higgs}(0) \approx -6 \times 10^{-5}$,
with even smaller values for larger Higgs boson masses.
\bigskip

\centerline{\bf REFERENCES}
\begin{enumerate}

\item[{[1]}] B.~W.~Lynn, M.~Peskin, and R.~G.~Stuart in {\em Tests of
Electroweak Theories:  Polarized Processes and Other Phenomena},
Proceedings of the 2nd Conference on Tests of Electroweak Theories,
Trieste, Italy, June 10-12, 1985, edited by B.~W.~ Lynn and C.~
Verzegnassi (World Scientific, Singapore, 1986), p.~213.

\item [{[2]}] M.~E.~Peskin and T.~Takeuchi, \PRL {\bf 65}, 964 (1990); Phys.
Rev. D {\bf 45}, 381 (1992).

\item[{[3]}] B.~Holdom and J.~Terning, \PLB {\bf 247}, 88 (1990); R. Johnson,
B.-L. Young, and D. W. McKay, \PRD {\bf 42}, 3855 (1990); {\bf 43}, R17 (1991);
M.~Golden and L.~Randall, \NPB {\bf 361}, 3 (1991); A. Dobado, D. Espriu,
and M. L. Herrero, \PLB {\bf 255}, 405 (1991).

\item[{[4]}] G.~Altarelli and R.~Barbieri, \PLB {\bf 253}, 161 (1991);
G.~Altarelli, R.~Barbieri, and S.~Jadach, \NPB {\bf 369}, 3 (1992).

\item[{[5]}] W.~J.~Marciano and J.~L.~Rosner, \PRL {\bf 65}, 2963 (1990); {\bf
68}, 898(E) (1992).

\item[{[6]}] D.~C.~Kennedy and P.~Langacker, \PRL {\bf 65}, 2967 (1990);
{\bf 66}, 395(E) (1990); \PRD {\bf 44}, 1591 (1991).

\item[{[7]}] P.~Sikivie, L.~Susskind, M.~Voloshin, and V.~Zakharov, \NPB
{\bf 173}, 189 (1980).

\item[{[8]}] CDF Collaboration, F.~Abe {\it et al.,} \PRD {\bf 43}, 2070
(1991).

\item[{[9]}] UA2 Collaboration, J.~Alitti {\it et al.,} \PLB {\bf 276}, 354
(1992).

\item[{[10]}] A.~Denner and T.~Sack, Z.~Phys.~C {\bf 46}, 653 (1990).

\item[{[11]}] A.~Denner, Fortschr.~Phys.~{\bf 41}, 307 (1993).

\item[{[12]}] I.~Maksymyk, C.~P.~Burgess, and D.~London, McGill Univ. report
93/13 and Univ. of Montreal report UdeM-LPN-TH-93-151, June, 1993
(unpublished).

\item[{[13]}] B.~Grinstein and M.~Wise, \PLB {\bf 265}, 326 (1991).

\item[{[14]}] By $\Pi_{XY}$ we mean the coefficient of $g_{\mu\nu}$ in the
vacuum polarization tensor. The $q^\mu q^\nu$ part can be neglected.  See
Ref.~[2] for details.

\item[{[15]}] The prime stands for differentiation with respect to $q^2$. This
is not to be confused with the notation of Ref.~[2], where $\Pi'_{XY}(q^2)$
stands for $[\Pi_{XY}(q^2) - \Pi_{XY}(0)]/q^2$.

\item[{[16]}] The Working Group on LEP Energy and the LEP Collaborations
ALEPH, DELPHI, L3 and OPAL, \PLB {\bf 307}, 187 (1993).

\item[{[17]}] D.~Albert, W.~J.~Marciano, D.~Wyler, and Z.~Parsa, Nucl.~Phys.
{\bf B166}, 460 (1980).

\item[{[18]}] D.~Yu.~Bardin, S.~Riemann, and T.~Riemann, Z.~Phys.~C {\bf 32},
121 (1986).

\item[{[19]}] A.~Sirlin, \PRD {\bf 22}, 971 (1980);
W.~Hollik, Fortschr.~Phys.~{\bf 38}, 165 (1990).

\item[{[20]}] S.~Bethke, in {\em Proceedings of the 26th International
Conference on High Energy Physics, Dallas, Texas, August, 1992}, edited by
J.~Sanford (World Scientific, Singapore, 1992).

\item[{[21]}] S.~Kopp, Thesis, University of Chicago, 1993 (CDF Collaboration);
F.~Abe {\it et al.}, paper submitted by the CDF Collaboration to XVI
International Symposium on Lepton and Photon Interactions, Cornell University,
Ithaca, NY, Aug.~11--15, 1993.

\item[{[22]}] D0 Collaboration, paper submitted to 1993 Cornell Symposium (see
Ref.~[21]).

\item[{[23]}] N. Cabibbo, in {\it Proceedings of the Third Topical Workshop on
Proton-Antiproton Collider Physics}, Rome, 12-14 January 1983, edited by C.
Bacci and G. Salvini (CERN, Geneva, 1983), p.~567;
F.~Halzen and K.~Mursala, \PRL {\bf 51}, 857 (1983).

\item[{[24]}] A.~D.~Martin, W.~J.~Stirling, and R.~G.~Roberts, \PLB {\bf 228},
149 (1989).  An updated set of structure functions by the same authors implies
a slightly higher ratio of $3.33 \pm 0.03$.  See A.~D.~Martin, W.~J.~Stirling,
and R.~G.~Roberts, Rutherford Appleton Laboratory report RAL-92-078, November,
1992 (unpublished).

\item[{[25]}] R. Hamberg, W. L. van Neerven, and T. Matsuura, \NPB
{\bf 359}, 343 (1991).

\item[{[26]}] M.~Swartz, invited talk at 1993 Cornell Symposium (see
Ref.~[20]).

\item[{[27]}] See, e.g., R.~Rosenfeld and J.~L.~Rosner, \PRD {\bf 38}, 1530
(1988).

\item[{[28]}] S. Bertolini, Nucl.~Phys. {\bf B272}, 77 (1986).

\item[{[29]}] A. Denner, J. Guth, and J. H. K\"uhn, \PLB {\bf 240}, 438 (1990).

\item[{[30]}] L. Lavoura and L.- F. Li, Carnegie-Mellon Report No.
CMU-HEP93-17, September 1993 (unpublished).

\item[{[31]}] CDF Collaboration, F.~Abe {\it et al.,} \PRL {\bf 67}, 2610
(1991).

\item[{[32]}] The importance of continuum lepton pair production in searching
for non-standard physics has been emphasized recently by W. T. Giele, E. W. N.
Glover, and D. A. Kosower, FERMILAB-Pub-93-086-T, April, 1993.

\item[{[33]}] We thank Henry Frisch for a conversation on this point.
\end{enumerate}
\newpage

\renewcommand{\arraystretch}{1.3}
\centerline{Table I.  Finite parts of contributions to $\delta_W^{\rm sm}(0)$.}
\bigskip
\begin{center}
\begin{tabular}{l c r} \hline
Contribution     & Coefficient of      & Value \\
                 & $g^2 Z_W/(4 \pi)^2$ & (\%)~ \\ \hline
Light fermions   &        3            & 0.80 \\
$t \bar b$ loop  & $-0.15^a$           & $-0.04^a$ \\
Photon - $W$     & $-1.00$             & $-0.27$ \\
$Z$ loops        & 0.51                & 0.14 \\
Standard Higgs   & $-0.02^b$           & $-0.006^b$ \\ \hline
Total            & $2.34$              & $0.62 $ \\ \hline
\end{tabular}
\end{center}
\centerline{$^a$For $m_t = 140$ GeV/$c^2$.~~~$^b$For $M_H = 100$ GeV/$c^2$.}
\bigskip
\bigskip

\centerline{Table II.  Finite parts of contributions to
$\delta_V^{\rm sm}$ and $\delta_\mu$.}

\bigskip
\begin{center}
\begin{tabular}{l c r c r} \hline
& \multicolumn{2}{c}{Leptons:} & \multicolumn{2}{c}{Quarks:} \\
Contribution   & Coefficient of & Value & Coefficient of & Value \\
               & $g^2 Z_W/(4 \pi)^2$ & (\%)~ & $g^2 Z_W/(4 \pi)^2$ &
(\%)~ \\ \hline
Wave function  & $-0.11$       & $-0.03$  & 0.28    & 0.07 \\
Vertices       & $-0.91$       & $-0.24$  & $-2.28$ & $-0.61$ \\
Bremsstrahlung & $-0.08$       & $-0.02$  & 0.72    & 0.19 \\
$\delta_\mu$   & $-2.55$       & $-0.68$  & $-2.55$ & $-0.68$ \\ \hline
Subtotal       & $-3.65$       & $-0.97$  & $-3.83$ & $-1.03$  \\ \hline
Total$^a$      & $-1.31$       & $-0.35$  & $-1.49$ & $-0.41$ \\ \hline
\end{tabular}
\end{center}
\centerline{$^a$Including contributions of Table I.}
\bigskip
\bigskip

\centerline{\bf FIGURE CAPTIONS}
\bigskip

\noindent
FIG.~1.  Correction term $\delta_W(0)$ affecting $W$ partial width to $e \nu$.
Solid line:  contribution from top quark as function of $m_t$; dashed line:
contribution from Higgs sector as function of charged Higgs boson mass $m_+$;
dotted line: contribution from extra degenerate lepton doublet as function of
mass $m_\ell$.
\end{document}